\documentclass[runningheads]{llncs}

\usepackage{amsfonts}
\usepackage{amssymb}
\usepackage{graphicx}
\usepackage[utf8x]{inputenc}
\usepackage[ruled]{algorithm2e}
\usepackage{amsmath}
\usepackage{xcolor}%

\usepackage{graphicx} 

\begin{document}

\title{Explainable Post hoc Portfolio Management Financial Policy of a Deep Reinforcement Learning agent}
\titlerunning{XDRL for Financial Portfolio Management}
\author{Alejandra de la Rica Escudero, Eduardo C. Garrido-Merchán, \\ María Coronado-Vaca}
\authorrunning{Alejandra de la Rica Escudero et al.}
\date{July 2024}

\institute{Universidad Pontificia Comillas, Madrid, Spain \\
\email{201906490@alu.comillas.edu, ecgarrido@comillas.edu, mcoronado@comillas.edu}}

\maketitle

\begin{abstract}
Financial portfolio management investment policies computed quantitatively by modern portfolio theory techniques like the Markowitz model rely on a set on assumptions that are not supported by data in high volatility markets such as the technological sector or cryptocurrencies. Hence, quantitative researchers are looking for alternative models to tackle this problem. Concretely, portfolio management is a problem that has been successfully addressed recently by Deep Reinforcement Learning (DRL) approaches. In particular, DRL algorithms train an agent by estimating the distribution of the expected reward of every action performed by an agent given any financial state in a simulator, also called gymnasium. However, these methods rely on Deep Neural Networks model to represent such a distribution, that although they are universal approximator models, capable of representing the previous distribution over time, they cannot explain its behaviour, given by a set of parameters that are not interpretable. Critically, financial investors policies require predictions to be interpretable, to assess whether they follow a reasonable behaviour, so DRL agents are not suited to follow a particular policy or explain their actions. In this work, driven by the motivation of making DRL explainable, we developed a novel Explainable Deep Reinforcement Learning (XDRL) approach for portfolio management, integrating the Proximal Policy Optimization (PPO) deep reinforcement learning algorithm with the model agnostic explainable machine learning techniques of feature importance, SHAP and LIME techniques to enhance transparency in prediction time. By executing our methodology, we can interpret in prediction time the actions of the agent to assess whether they follow the requisites of an investment policy or to assess the risk of following the agent suggestions. To the best of our knowledge, our proposed approach is the first explainable post hoc portfolio management financial policy of a DRL agent. We empirically illustrate our methodology by successfully identifying key features influencing investment decisions, which demonstrate the ability to explain the agent actions in prediction time. 
\end{abstract}


\section{Introduction}
Financial portfolio management is a critical task in the investment domain \cite{drake2010basics}, traditionally guided by modern portfolio theory techniques such as the Markowitz model \cite{mangram2013simplified}. These models, which optimize the trade-off between risk and return, are grounded in a set of assumptions about market behavior \cite{wilford2012true}. However, for example, in high-volatility markets like the technological sector or cryptocurrencies, these assumptions often fail to hold true \cite{hill2013different}. This discrepancy has led quantitative researchers to seek alternative methods that can better handle the dynamic and unpredictable nature of such markets \cite{rather2017stock}. 

The application of Deep Reinforcement Learning (DRL) \cite{arulkumaran2017deep} to portfolio management has gained popularity in recent years.
Concretely, DRL algorithms train agents to maximize expected returns by learning optimal actions through interaction with a simulated environment \cite{sutton2020reinforcement}, also known as gymnasium \cite{brockman2016openai}. These agents use Deep Neural Networks models (DNNs) \cite{lecun2015deep} to approximate the distribution of expected rewards for different actions in varying financial states. Despite their capability as universal function approximators \cite{lu2020universal}, DNNs suffer as they lack interpretability \cite{castelvecchi2016can}. This means a challenge in financial contexts, where decision-making transparency is crucial for investors to trust and adopt automated strategies \cite{lowenstein1996financial}.

The need for explainability in financial decision-making drives the development of Explainable Artificial Intelligence (XAI) techniques \cite{caruana2020intelligible,angelov2021explainable,mehta2023explainable}. In the context of DRL, incorporating XAI methods enhance the high performance of DRL agents and the necessity for transparent, interpretable investment policies. But, as we will show in section 2 (sate of the art), despite the growing literature that analyzes the application of DRL to portfolio management, the literature on the explainability of DRL algorithms applied to  portfolio management is very scarce and underdeveloped, with only four recent studies \cite{wang2019alphastock,bougie2020towards,guan2021explainable,shi2021xpm}, to the best of our knowledge. Moreover, these four published DRL explainability methods in portfolio management, only offer explanations of the model in training time, not being able to monitor the predictions done by the agent in the trading time.

Driven by this motivation, and to respond to this research gap in the literature, in our work, we introduce a novel Explainable Deep Reinforcement Learning (XDRL) framework for portfolio management. Concretely, our approach combines the popular Proximal Policy Optimization (PPO) algorithm \cite{schulman2017proximal}, a state-of-the-art DRL technique, with model-agnostic XAI methods such as feature importance \cite{belle2021principles}, SHAP (SHapley Additive exPlanations) \cite{rozemberczki2022shapley}, and LIME (Local Interpretable Model-agnostic Explanations) \cite{visani2022statistical}. By doing so, we obtain interpretability of DRL trained agents. The three explainability techniques can be implemented independently or jointly, being able to explain the DRL agent predictions in trading time, being able to track throughout the time whether the policy is acting as it is expected or not, what is an advantage with respect to the rest of four published DRL explainability methods just mentioned. Our working hypothesis is that the predictions of DRL agents can be explained in a post-hoc fashion, offering an interpretation that can be tested with respect to financial investment policies. 

Our paper contributes to the underdeveloped literature on the applications of XDRL models
in the realm of portfolio management in two ways: We add to this emerging literature of only four previous studies but with promising results and, our research novelty entails, to the best of our knowledge, the first-ever application of an explainable post-hoc portfolio management policy of a DRL agent. 

The rest of the paper is organised as follows. First, we begin with a state-of-the-art section where we will show the work that is currently done on DRL and XDRL in financial portfolio management. Afterwards, we will continue with a methodology section where we will explain the fundamental details of deep reinforcement learning applied to financial portfolio management, the PPO algorithm and the explainability techniques that are going to be used to explain the predictions of the agents. We continue our work with an experiments section where we give empirical evidence about the usefulness of our approach that supports our claim that DRL predictions can be explained and hence compared with a particular financial policy. Finally, we illustrate conclusions and further research lines in this useful and significant topic of explaining DRL agent behaviour to obtain transparency about the financial investment policy that is being executed by an agent.

\section{State of the art}
The application of DRL in financial portfolio management is gaining popularity in the recent years, mainly due to the rise of computing power and architectures that enables a reasonable estimation of the rewards distribution of the actions with respect to the states given by the training process of the agents with respect to financial data \cite{hambly2023recent}. But despite the growing literature that analyzes the application of DRL to portfolio management, the literature on the explainability of DRL algorithms applied to  portfolio management is very scarce and underdeveloped, with only four recent studies \cite{wang2019alphastock,bougie2020towards,guan2021explainable,shi2021xpm}, to the best of our knowledge. In this section, we show a detailed state-of-the-art of DRL and XDRL applied to financial portfolio management to show the research gap in the literature to which our work responds.  

Multiple DRL algorithms have been proposed recently, which has motivated their application in our area of interest. Liu et al. \cite{liu2021finrl} classify the state-of-art DRL algorithms into three categories: 1) value-based algorithms: those based on Deep Q-Networks (DQN) \cite{mnih2015human}; 2) policy-based algorithms: directly update the parameters of a
policy through policy gradient (PG) \cite{sutton1999policy}; and 3) Actor-Critic based algorithms, such as Advantage Actor Critic (A2C) \cite{mnih2015human}, Proximal Policy Optimization (PPO) \cite{schulman2017proximal}, Deep Deterministic Policy Gradient (DDPG) \cite{lillicrap2015continuous}, Soft Actor-Critic (SAC) \cite{haarnoja2018soft}, or Twin Delayed Deep Deterministic Policy Gradient (TD3) \cite{dankwa2019twin}. And all of them 
have been used to maximize portfolio returns while minimizing risk; for example, \cite{liu2018practical} apply DDPG for stock trading; \cite{liu2020finrl} apply DQN, DDPG, PPO, A2C, TD3, SAC for automated stock trading in quantitative finance; \cite{liu2021finrl} implement PPO, A2C, DDPG and TD3 for portfolio management. Liang et al. \cite{liang2018adversarial} have applied successfully to financial portfolio management three state-of-the-art popular DRL algorithms that can deal with continuous valued actions, namely PPO, DDPG and PG. Specifically, the authors conducted comprehensive experiments on the China stock market, examining various settings, including learning rates, objective functions, and feature combinations, to derive insights for parameter tuning, feature selection, and data preparation. Additionally, driven by the usefulness of DRL in high volatility markets, some authors have applied this methodology in cryptocurrency portfolio management \cite{liang2018adversarial,jiang2017cryptocurrency,sadighian2019deep,sattarov2020recommending,liu2021finrl}. Zhang et al. \cite{zhang2019deep} apply DRL to trade futures contracts and various authors implement hedging strategies with DRL (deep hedging) \cite{buehler2019deep,cao2021deep,carbonneau2021deep}. More concretely, Pham et al. \cite{pham2021multi} focus on multi-agent DRL to automatically construct hedging strategies. And other authors also apply multi-agent DRL in portfolio management \cite{lussange2021modelling,huang2022mspm,lee2020maps}. \cite{zhang2020cost} propose a cost-sensitive portfolio selection with DRL, being, thus, the first authors to incorporate transaction costs. While all these DRL approches for portfolio management only consider price changes of assets, but without considering the relation between companies, \cite{shi2022gpm} propose a new DRL framework for portfolio management based on GCN (Graph Convolutional Network) to take into account the relational features of the portfolio (relations between assets and their corresponding companies). \cite{hao2023stock} incorporate ensemble techniques and fuzzy extension in addition to existing DRL algorithms and use them for portfolio management.  \cite{jang2023deep} propose a novel DRL approach for portfolio optimization that combines the MPT and a DL approach (specifically, they solve the multimodal problem on a dataset of 28 USA stocks through the Tucker decomposition of a model with the input of technical analysis and stock return covariates). Some authors incorporate sentiment analysis in DRL to also perceive market sentiment in portfolio allocation \cite{koratamaddi2021market,li2019optimistic,nousi2023leveraging,azhikodan2019stock}.     Hambly et al. \cite{hambly2023recent} provide a review of the recent developments and use of RL and DRL in finance, including portfolio management. 
 
But despite the growing literature that analyzes the application of DRL to portfolio management (as we have just reviewed above), the literature on the explainability of DRL algorithms applied to  portfolio management is very scarce and underdeveloped, with only four recent studies \cite{wang2019alphastock,bougie2020towards,guan2021explainable,shi2021xpm}, to the best of our knowledge. Guan and Liu (2021) \cite{guan2021explainable} provide an empirical approach of explainable DRL for the portfolio management task in response to the challenge of understanding a DRL-based trading strategy because of the black-box nature of deep neural networks. Specifically, they use a linear layer in hindsight as the reference model and they find the relationship between the reward (the portfolio return) and the input (the features) by using integrated gradients. Since the linear model (a regression) is interpretable given that the coefficients are interpretable, then the methodology is interpretable. The neural network's capacity as a universal approximator dramatically exceeds that of regression. However, they use a neural network and then make a kind of compensation between the network and the coefficients to see how the linear regression "approximates" the network. If it approximates well, then you can trust of the coefficients. But you lose explainability if the approximation is poor. Additionally, as they are using linear interpretation, correlations may not explain complex patterns. Moreover, their approach differs from ours since it is an explainability approach dependent on the model, while ours is agnostic of the model, it is a post-hoc one. Bougie and Ichise (2020) \cite{bougie2020towards} present a method to combine external knowledge and interpretable reinforcement learning in portfolio management. They derive a rule-based variant version of the Sarsa algorithm (\cite{sutton2020reinforcement}, p.140), that is, a neurosymbolic. This way, you can thus add a priori rules and data augmentation to "explain" your policy, since you are "injecting" it a priori. It is not a post-hoc approach like ours but rather an a priori one. While we explain the agent's predictions, they inject the agent with a policy in the form of rules before training. In fact, as we state in section  5 (Conclusions and further research) a line of future work could be to hybridize their approach with ours and see how training modifies the rules injected a priori. Shi et al. (2021) \cite{shi2021xpm} add explainability to their DRL methods in portfolio managememt through this approach: they use a temporal neural network to extract significant features that explain the patterns as a function of time, that is, a model that manages time, compared to our CNN. Then they apply a regularization technique to simplify them and finally, to explain them, they use class activation mapping (CAM), a way to explain the features of the neural network that they have used in the model, not in prediction time, like our proposed approach.  Thus, they explain the model, that is, the policy trained during the training period, but we explain the predictions of the model during trading time. Wang et al. (2019) \cite{wang2019alphastock} offer an interpretable DRL investment strategy using interpretable deep attention networks. A ranking of features is obtained through two neural networks that seem to explain the training time data in the best possible way and subsequently a sensitivity analysis is performed to determine the best ones. Their interpretation analysis results reveal that their strategy selects assets by following a principle as “selecting the stocks as winners with high long-term growth, low volatility, high intrinsic value, and being undervalued recently”. Once again, this approach differs from ours since it is useful for explanatory purposes, while we make explanations of predictions, for predictive purposes. 

Thus, to the best of our knowledge, our study is the first to propose an explainable post hoc portfolio management financial policy of a Deep Reinforcement Learning agent. The other only four existing studies in this underdeveloped strand of literature offer an explainable "a priori" investment strategy using DRL models.

\section{Methodology}
We now introduce the methodological details of our proposed approach to post-hoc explainable deep reinforcement learning applied to financial portfolio management. First, we will explain the fundamentals of deep reinforcement learning applied to finance, then, we will illustrate the explainable artificial techniques that we have chosen and, finally, we will show how we can integrate those techniques into the deep reinforcement learning method to explain the predictions of the agent.   

\subsection{Fundamentals of Deep Reinforcement Learning applied to financial portfolio management}
We will first introduce objections to our methodology for portfolio management and arguments that answer to those objections. Then, we will describe the fundamentals of deep reinforcement learning and how we can apply these algorithms to financial portfolio management.

Although DRL has potential for portfolio management due to its competence in capturing nonlinear features, low prior assumptions, and high similarities with human investing, there are characteristics worth paying attention to as pointed out by Liang et al. \cite{liang2018adversarial}: First, a financial market is both highly volatile and non-stationary, totally different to games or robot control \cite{liu2021deep,shao2019survey} which are the main sectors where DRL has been experimented. Second, traditional Reinforcement Learning (RL) aims to maximize rewards over an infinite period, while portfolio management focuses on maximizing returns within a finite time. Third, in finance, it's crucial to test strategies on separate data sets to evaluate their performance, unlike in games or robotics. Lastly, the stock market has an explicit expression for portfolio value; therefore, approximating the value function is useless and can even deteriorate the agent´s performance. 

However, deep neural networks are able to approximate any function given enough data and a particular architecture of the network, being universal approximator functions. Regarding maximizing returns within a finite time, we can tune the DRL algorithm via the $\gamma$ hyperparameter to consider high future rewards. Concretely, $\gamma \in [0,1]$ controls the focus of the agent in immediate or far rewards as a function of time where a value near to one focus on maximizing returns on a long time period, being $\gamma$ the same as a discount rate for all DRL algorithms. Next, we can assume that an immediate future behaviour of the stock market is explained technically and by past information, being also DRL suited in this scenario. The agent can be retrained in a constant fashion after its predictions happen in the real-time scenario with the new information. Lastly, although Markowitz model assumes that portfolios are only a function of expected reward and risk, if those assumptions, like normal distributed returns, are not met, then, the function explaining the optimal portfolio is a black-box of an enormous set of features like technical indicators, fundamental ratios or social networks, that DRL algorithms can handle due to neural scaling laws. In this work, we assume that the market can be perfectly explained by technical data, focusing hence only in this kind of data. However, any source of data can also be integrated into the space state of the agent, even multimodal data, so this assumption is not an issue in real-case scenarios. To sum up, we consider that DRL can be successfully applied to the financial portfolio management problem, and now explain the fundamental concepts of this methodology.

DRL is a class of methods that combines reinforcement learning algorithms \cite{sutton2020reinforcement,hambly2023recent} with deep neural network models \cite{lecun2015deep} to tackle any complex decision-making task. In DRL, an agent interacts with an environment defined by a state space $\mathcal{S}$ and an action space $\mathcal{A}$. Critically, these spaces can be a bounded continuous domain, $\mathcal{S} \in \mathbb{R}^d$ and $\mathcal{A} \in \mathbb{R}^d$ where $d$ is the number of features that the agent perceives in each time step $t$, and not limited to discrete spaces, as in reinforcement learning. In particular, deep reinforcement learning will encode the policy $\pi(a_t|s_t)$ learnt by trail and error with the environment in the training process in the deep neural network that will map a distribution of states to actions $\mathcal{S} \to \mathcal{A}$, with the purpose of selecting the action that maximizes an expected reward in a given time period by a $\gamma \in [0,1]$ hyperparameter. Concretely, at each discrete time step $t$, the agent observes a state $s_t \in \mathcal{S}$ and selects an action $a_t \in \mathcal{A}$ based on the learnt policy $\pi(a_t|s_t)$, that acts as the conditional probability distribution of actions given states being estimated to maximize the expected reward. The environment responds to the action by transitioning to a new state $s_{t+1}$ and providing a reward $r_t$ as an effect of making action $a_t$ given state $s_t$. Following an iterative process in a simulator, the agent can learn a policy $\pi(a_t|s_t)$ by trail and error that may generalize outside of the simulator if the assumptions made by the deep neural network model are met by the prediction data.

\begin{figure}[h]
  \centering
  \includegraphics[width=0.99\textwidth]{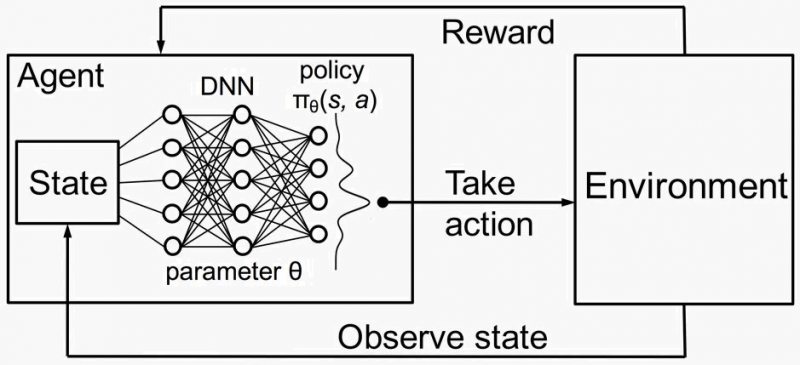} 
  \caption{Deep reinforcement learning main components that enable the estimation of the expected reward of any action of the action space of the agent conditioned to any state perceived by the agent. The learnt policy function is encoded by a deep neural network, enabling continuous-valued actions and spaces and any complexity of its mapping.}
  \label{fig:dnn}
\end{figure}

More formally, the purpose of the learning process is to learn a policy $\pi$ that maximizes the expected cumulative reward, which can be defined with a return function in every time step $R_t$:

\begin{equation}
    R_t = \sum_{k=0}^{K} \gamma^k r_{t+k}\,,
\end{equation}

where $\gamma \in [0,1]$ is the previously mentioned discount factor that balances immediate and future rewards and $K$ is the end of the episode, or desired prediction period. Consequently, once that we have estimated the policy that maximizes the expected return function $R_t$, that can be personalized for any problem, we can define, for every time step and for every action and state, a q-value function $Q(s,a)$ that represents the complete distribution of the expected return of taking any action $a \in \mathcal{A}$ in any state $s \in \mathcal{S}$ and following policy $\pi(a_t|s_t)$ as:

\begin{equation}
    Q^{\pi}(s,a) = \mathbb{E}_{\pi} [R_t | s_t = s, a_t = a]\,,
\end{equation}

that is the probability distribution learnt by a DRL algorithm and approximated in the used deep neural network $Q(s,a|\theta)$ by its set of parameters $\theta$. We illustrate this framework in Figure \ref{fig:dnn}.

One of these algorithms, used in our work as its popularity and being a lightweight version of a more expensive algorithm, is Proximal Policy Optimization (PPO) \cite{schulman2017proximal}, that aims to update policies without making big changes at once to avoid outliers in the training process that can make the estimation of the policy worse. This makes the learning process more stable. This is achieved by creating a "trust region," ensuring that new policies don't deviate too much from old ones. In order to do so, PPO introduces a clipped objective function (clipped probability ratios) that prevents the updating step from moving the new policy too far from the old policy. This clipping mechanism modifies the policy optimization by clipping the probability ratio between the new and old policies, keeping it within a specified range. By maximizing the clipped objective, PPO finds a balance between trying new strategies (exploration) and sticking with known good ones (exploitation).

\subsection{Explainable artificial intelligence techniques}
As the purpose of our work is to enhance the DRL framework by integrating explainability of financial features to make the decision-making process of the DRL model transparent and understandable, we briefly describe in this section some of the techniques that we have integrated in the DRL framework to make it interpretable for financial experts.

Human decision makers use predictions made by ML models in their process, but the usability of those predictions is limited if the human is unable to justify and understand their trust in said predictions. Explanation is a way of obtaining such understanding, by selecting “what” must be communicated and “how” that information is presented. Moreover,according to \cite{adadi2018peeking} XAI is necessary
for regulatory issues, and thus, there is also a legal component to be considered. The EU General Data Protection Regulation (GDPR) aims to ensure the ‘right to explanation’ concerning automated decision-making models. 

Interpretability also helps developers understand and improve the model performance. It can aid in troubleshooting and debugging, as well as in detecting potential biases in the AI system and provide insights into how the system would react under different circumstances. This understanding can also provide insights that go beyond predictions, uncovering underlying patterns and relationships within the data.

Vouros \cite{vouros2022explainable} provides a comprehensive review of state-of-the-art  for explainable DRL methods, categorizing them in classes according to the paradigm they follow, the interpretable models they use, and the surface representation of explanations provided. In fact, it is the unique existing literature review focusing on XDRL.  
In our work, we have used three explainability techniques: the SHapley Additive exPlanations (SHAP) \cite{lundberg2017unified}, the Local Interpretable Model-agnostic Explanations (LIME) methodology \cite{ribeiro2016should}, and feature importance methods. 

The SHAP method offers a unified way to explain each instance´s predictions \cite{lundberg2017unified}. Imagine you're playing a team sport, and at the end, you want to know who contributed most to the win. It's not enough to say "everyone did their best"; you want specifics, like who scored the most goals. SHAP does this for machine learning. It breaks down a prediction to show the impact of each feature—like how a player's actions affect the game's outcome.
Here is how SHAP works in very simple terms:
1.	Contribution: It looks at each feature (like a player in a game) of the data the model uses and asks, "What’s the contribution of this feature to the final prediction?".
2.	Fair Distribution: Then, SHAP uses a fair method for distributing the "credit" of the outcome. It makes sure that the contributions of all features sum up to the total prediction. This is like making sure that all the individual scores from players add up to the final score of the game.
3.	Individual Impact: SHAP values can tell you the impact of a single feature on the prediction. For example, just as you might wonder how much scoring that one goal early on helped the team win, SHAP can show how much changing one piece of data can change the prediction.
4.	Teamwork: It also considers how features affect the prediction when they work together, kind of like how players might pass the ball to each other.
5.	Comparisons: It even lets you compare the importance of different features. Like after a match, you might debate who the most valuable player was, based on their contributions.

LIME is a tool designed to explain the predictions of any classifier in a way that is understandable and accurate \cite{ribeiro2016should}. It works by learning a simplified model around the prediction, using perturbations of uniformly sampled features. This allows for a better understanding of the relationship between input features and model responses. The goal is to ensure local fidelity, explaining how the model behaves around the specific instance being predicted.
While LIME was originally designed to inspect models using binary vectors, for instance representation, it also offers solutions for outcome explanation, as it highlights the importance of features in a local context only.
LIME is a universal method that enhances both local and global model interpretability. In the context of DRL, it can be used to explain any models, such as local decisions for action selection based on interpretable state feature representation, even if it wasn't specifically designed for that task. Functions can provide binary vectors, for instance representation, while any interpretable model can theoretically be used for outcome explanation, like a linear model or a decision tree. Visual means are used to provide surface representations of the instance's interpretable representations for model inspection, as well as explanation logic for local and model explainability. Although Ludenberg et al. \cite{lundberg2017unified} showed that LIME is a subset of SHAP and that SHAP outperformed LIME, LIME is still a useful tool, since it’s faster compared to SHAP and thus, LIME can be practical in cases where efficiency is important.

Feature importance methods provide insights into which features are most influential in the model’s overall decision-making process. By analyzing the importance of each feature, we can understand how the model prioritizes different aspects of the input data when making predictions. This helps identify key drivers of the model's behavior, enhancing transparency and trust in the model's decisions.

\subsection{Integrating explainability in deep reinforcement learning financial predictions}
In this section we describe how we integrate the explainability techniques mentioned in the previous subsection with the DRL methodology illustrated before. Concretely, our work builds on the existing framework from the GitHub repository "Reinforcement learning in portfolio management" (DeepCrypto, 2018)  \url{https://github.com/deepcrypto/Reinforcement-learning-in-portfolio-management-/tree/master?tab=readme-ov-file}. The goal is to enhance this framework by integrating explainability features to make the decision-making process of the DRL model transparent and understandable. 

Our starting point is the already developed PPO-based model in the framework, which has demonstrated effectiveness in portfolio management tasks. The PPO algorithm is chosen for its balance between exploration and exploitation, making it well-suited for dynamic and unpredictable environments like financial markets.

We design a training phase consisting on financial data downloaded from Yahoo Finance with OHCLV (Open, High, Close,
Low, and Volume) information about several tickers to build a portfolio in a certain period of time. Once the agent is ready, after a preprocessing phase to ensure the data is clean and ready for analysis. This involves normalizing values, handling missing data and structuring the data so that the model can work with it. Then, the training phase starts on the financial market loaded, the agent interacts with the environment, learns from the data, and refines its policy to improve decision-making. 

To implement the explainability techniques in the already analyzed framework, we will use post-hoc interpretability methods. This requires saving the state-action pairs for all steps during the training process. These state-action pairs will be used later to create explainability models. To achieve our objective, we implement explainability techniques such as SHAP, LIME, and feature important methods. These methods reveal which features and inputs are most influential in the agent decision-making process, providing insights both at a global level and for specific predictions, leading to an interpretable DRL model.

\section{Experiments}
In this section, we will describe the different experiments that we have performed to show the usefulness of our explainable deep reinforcement learning methodology. For reproducibility and transparency, we have uploaded all the code of these experiments in the following Github repository \url{https://github.com/aleedelarica/XDRL-for-finance}.

For our portfolio management experiment we have considered a training period from 2015 to 2017 and a trading period from 2017 to 2018. We have considered the OHCL information of five different technological assets, having a total of $20$ features in the state space of the agent. We have considered a PPO DRL algorithm to learn the weights of the deep neural network with default hyperparameters and 100 epochs across all the financial data downloaded from Yahoo finance. Having all that information and performing the training period of the neural network, we interpret the predictions of the deep neural network in the trading period using feature importance, SHAP and LIME methods as we will describe in this section. 

\begin{figure}[h!]
  \centering
  \includegraphics[width=0.99\textwidth]{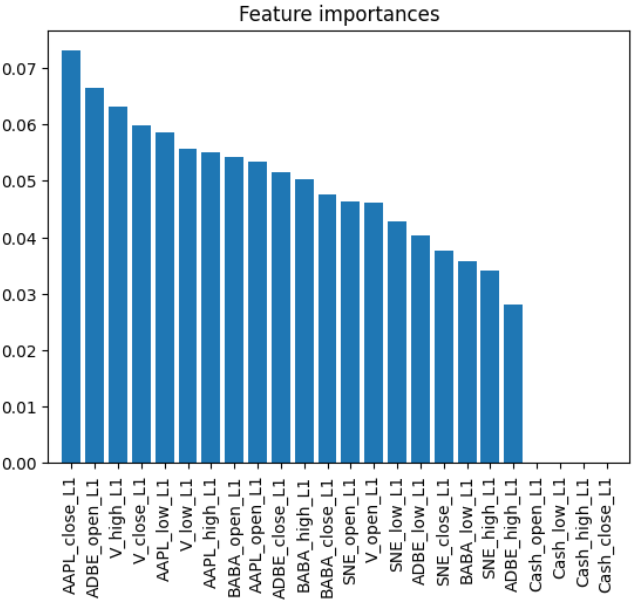} 
  \caption{Importance of the features used as the state space of the DRL agent for financial portfolio management experiments. We can see how the APPLE close value is the most important for the estimated policy of the DRL agent. }
  \label{fig:exp_1}
\end{figure}

We begin with experiments that show how the feature importance can be extracted for the predictions of the agent during the trading period. In order to do so, we configure a portfolio of several technological assets and measure the importance of their OHCL features in the trading period, information that we illustrate in Figure \ref{fig:exp_1}. We will then illustrate how can SHAP and LIME methods offer interpretability and explainability of the actions performed by the agent in such a scenario.

\begin{figure}[h]
  \centering
  \includegraphics[width=0.99\textwidth]{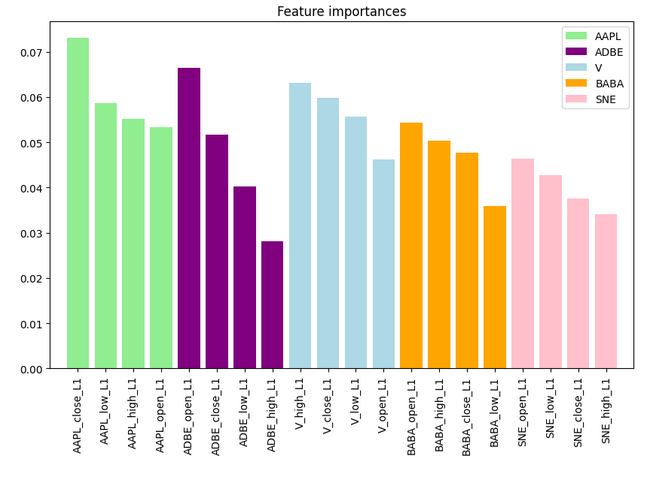} 
  \caption{Feature importance of the state space of the DRL agent sorted by assets.}
  \label{fig:exp_2}
\end{figure}

Figure \ref{fig:exp_2} shows the average importance across all features for each stock. Apple (AAPL) remains the most significant, followed by Visa (V), Alibaba (BABA), Adobe (ADBE), and Sony (SNE). This means the DRL agent in the model prioritizes Apple’s data in order to make the investing decisions, as it is consistently seen in the previous experiment. Most critically, we can see how, for every asset, it is not clear which of the OHCL is the most important one, as it varies in every asset. Consequently, it is wise to use them all to make the agent more robust to different portfolios. 

\begin{figure}[h]
  \centering
  \includegraphics[width=0.99\textwidth]{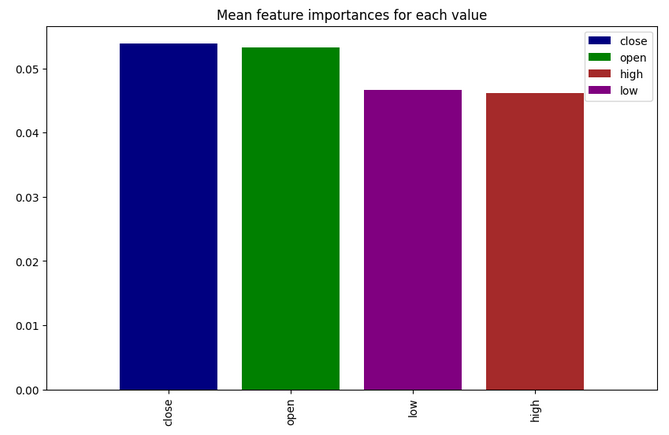} 
  \caption{Mean feature importance of the different financial indicators across all the assets of the portfolio.}
  \label{fig:exp_3}
\end{figure}

However, as we believe that it is important to know whether the OCHL information is important, we have made a plot of the average importance of each technical indicator for all the assets of the portfolio during all the trading period, obtaining the results plotted on Figure \ref{fig:exp_3}, where we can clearly see that although close and open are the most critical indicators, the extreme results of the day, high and low, are features that have as well importance in the predictions of the DRL agent.

To explain the model's predictions with SHAP, we use SHAP's force plot. Our model is a multi-output model (6 actions, one for each asset and 1 for the cash risk-free asset), consequently, SHAP can generate one force plot for each asset weight allocation through all the samples. We can observe the force plot for Apple’s weight allocation in Figure 5. 

\begin{figure}[h]
  \centering
  \includegraphics[width=0.99\textwidth]{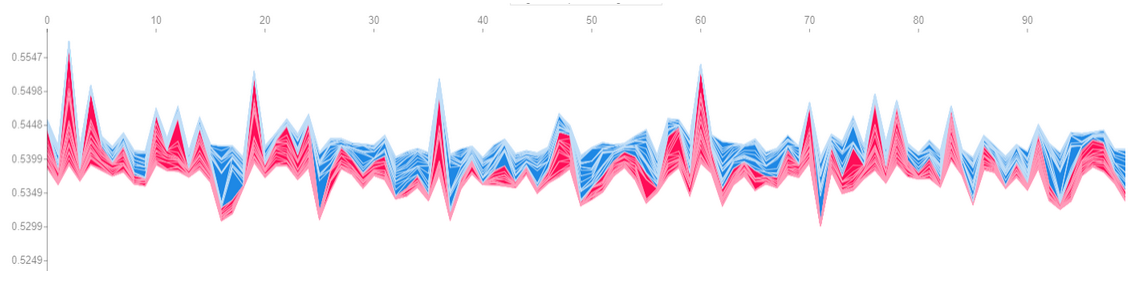} 
  \caption{SHAP force plot for AAPL weight allocation with all features contribution.}
  \label{fig:exp_4}
\end{figure}

\begin{figure}[h]
  \centering
  \includegraphics[width=0.99\textwidth]{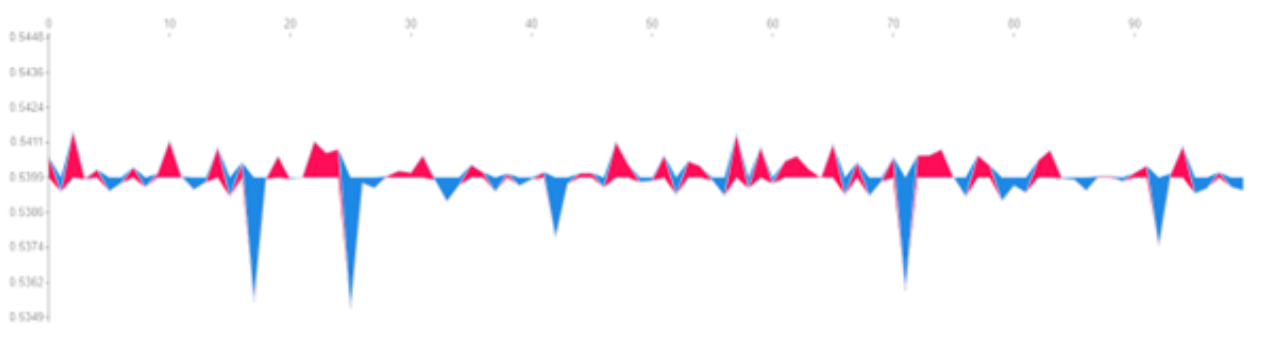} 
  \caption{SHAP force plot for AAPL weight allocation with only Apple's closing price value contribution.}
  \label{fig:exp_5}
\end{figure}

The SHAP values represent the contribution of each feature to the model's output for the specific sample. Positive SHAP values (red) push the prediction higher, while negative SHAP values (blue) push the prediction lower. The base value, represented by the horizontal baseline, is the average model output (logit in this case) over the entire dataset. Conversely, we can do the same process with only one feature, to test how and whether it impacts on the portfolio, as we illustrate in Figure \ref{fig:exp_5}, where we can see how the red and blue segments still indicate positive and negative contributions, respectively, but they are now isolated into a single feature (Apple’s closing price
value contribution, in this case). 

This plot helps in understanding both the specific impact of this particular feature through the entirety of the dataset and in one specific prediction without the interference of others. For instance, imagine an investor who wants to understand why the model lowered the weight allocation of Apple on a certain instance. The investor could analyze the impact of the different features to that prediction, realizing that the closing price of Apple stock lowered significantly this weight allocation prediction. The combination of the investor’s knowledge and the model´s explanation gives a complete understanding of the prediction. Continuing with the storyline, let's say the investor also examines another instance where the model increased the weight allocation of Apple. By analyzing this specific prediction, the investor notices a red segment indicating that the BABA\_open\_L1 value had a positive impact. This might initially seem puzzling, as BABA\_open\_L1 refers to the opening price of Alibaba stock from the previous day. However, the investor recalls that on this particular day, positive news about Alibaba's strong quarterly performance had a ripple effect on the tech sector, boosting overall market sentiment and indirectly benefiting Apple's stock price. The model's sensitivity to such interconnected market dynamics is reassuring to the investor. It demonstrates that the model doesn't just consider isolated stock movements but also understands broader market trends and their impacts on individual assets. This interconnected understanding is crucial for making informed allocation decisions in a diversified portfolio. This comprehensive explainability method enhances the investor's trust in the model, ensuring they can confidently rely on its predictions to guide their investment strategies. 

\begin{figure}[h]
  \centering
  \includegraphics[width=0.99\textwidth]{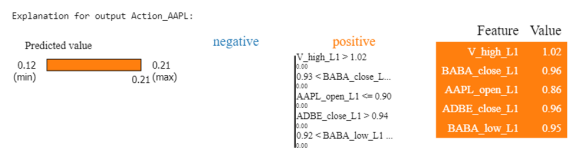} 
  \caption{LIME explanation for Apple's weight allocation prediction at a particular instance.}
  \label{fig:exp_6}
\end{figure}

Finally, we comment how can the LIME method be used to interpret the predictions of the DRL agent. By using LIME explanations, we can gain insights into the feature contributions for each asset's weight allocation at a specific instance, as we can see in the example illustrated on Figure \ref{fig:exp_6}.

The predicted weight allocation is at its maximum of 0.21. This high allocation is driven by several features.The positive contributors include V\_high\_L1, BABA\_close\_L1, AAPL\_open\_L1, ADBE\_close\_L1, and BABA\_low\_L1. These features collectively push the prediction higher. Specifically, the feature V\_high\_L1 with a value of 1.02 and BABA\_close\_L1 at 0.96 significantly influence the model's decision to allocate a larger portion to Apple. Notably, there are no significant negative contributors, indicating a strong overall positive sentiment for Apple’s allocation based on these metrics.

\begin{figure}[h]
  \centering
  \includegraphics[width=0.99\textwidth]{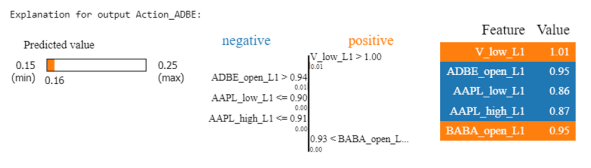} 
  \caption{LIME explanation for Adobe's weight allocation prediction at a particular instance.}
  \label{fig:exp_7}
\end{figure}

Moving to Adobe (Fig 8), the predicted allocation is 0.16 within a range of 0.15 to 0.25. Positive contributions come from V\_low\_L1 and BABA\_open\_L1. These metrics indicate that lower values for Visa and opening values for Alibaba push the allocation towards Adobe higher. On the negative side, for instance metrics from Apple, particularly AAPL\_low\_L1 and AAPL\_high\_L1, reduce the allocation. This insight suggests that while Adobe benefits from positive conditions in Visa and Alibaba, certain Apple metrics dampen its allocation.

This LIME explanations can assist investors by offering explicit explanations for each
forecast, allowing them to observe how different factors interact and influence the DRL
agent decisions. For example, if an investor finds that specific measures from Visa and
Alibaba continuously increase allocations to tech stocks such as Apple and Adobe, they may opt to constantly monitor these metrics. This actionable insight can help investors make more informed and proactive decisions.

Understanding why the model cuts allocations in response to specific traits can also assist investors in reducing risks. If Apple's poor metrics routinely reduce allocations to other stocks, the investor may investigate this link further and change their portfolio to balance potential drawbacks.

As we have shown throughout the section, the three explainability techniques can be implemented independently or jointly, being able to explain the DRL agent predictions in trading time, being able to track throughout the time whether the policy is acting as it is expected or not, what is an advantage with respect to the rest of published DRL explainability methods, that only offer explanations of the model in training time, not being able to monitor the predictions done by the agent in the trading time.  

\section{Conclusions and Further Work}

In this work, we successfully used SHAP, LIME and feature importance to explain the decisions made by our DRL model in portfolio management tasks. These methods provided clear and detailed insights into why the model made specific investment choices, affirming our hypothesis that it is possible to explain DRL predictions in portfolio management with explainability techniques. Also, our prediction explainability analysis with SHAP and LIME revealed that the importance of features varied across different market conditions and specific predictions. Moreover, our feature importance analysis showed that certain features consistently influenced the model's investment decisions more than others. 

We believe that several future directions could enhance and expand the impact of this work. First, it would be interesting to conduct several studies to evaluate how real investors interpret and react to the explanations provided by the model, which could yield valuable reinforcement learning from human feedback. For a better understanding of non-technical investors, it would be nice to develop an intuitive user interface that presents the model's decisions and their explanations in a user-friendly manner would improve its practicality. Also, it would be interesting to deal with a wider set of explainable techniques and different markets to assess the usefulness of the methodology in different scenarios. Finally, a future line of work would be to hybridize one of the a priori XDRL approaches of Bougie and Ichise \cite{bougie2020towards} or Shi et al. \cite{shi2021xpm} with our proposed post hoc approach (the only one in the literature of XDRL for portfolio management); that is, explaining both the model (the policy trained during the training period) as these authors do, and the predictions of the model during trading time (as we do), and see how training modifies the rules injected a priori \cite{bougie2020towards} or the policy trained \cite{shi2021xpm}.

\bibliographystyle{unsrt}

\bibliography{main}

\end{document}